\documentclass[aps,floats]{revtex4}
\usepackage{amsmath,amssymb}
\usepackage{graphicx,epsfig}

\begin{document}
\bibliographystyle {plain}

\def\oppropto{\mathop{\propto}} 
\def\opsimeq{\mathop{\simeq}}
\def\opoverderline{\mathop{\overline}}
\def\operarrow{\mathop{\longrightarrow}}
\def\opsim{\mathop{\sim}}

\def\fig#1#2{\includegraphics[height=#1]{#2}}
\def\figx#1#2{\includegraphics[width=#1]{#2}}


\title{ Non-equilibrium dynamics of polymers and interfaces in random media : 
\\ conjecture $\psi=d_s/2$ for the barrier exponent   } 


 \author{ C\'ecile Monthus and Thomas Garel }
  \affiliation{Service de Physique Th\'{e}orique, CEA/DSM/SPhT\\
 Unit\'e de recherche associ\'ee au CNRS\\
 91191 Gif-sur-Yvette cedex, France}

\begin{abstract}

We consider various random models (directed polymer, random ferromagnets, spin-glasses) 
in their disorder-dominated phases, where the free-energy cost $F(L)$ of
an excitation of length $L$ presents fluctuations 
that grow as a power-law $\Delta F(L) \sim L^{\theta}$
with the 'droplet' exponent $\theta$.
Within the droplet theory, the energy and entropy 
of such excitations 
present fluctuations that grow as
$\Delta E(L) \sim \Delta S(L) \sim L^{d_s/2}$ where $d_s$ is the dimension
of the surface of the excitation.
These systems usually present a positive 'chaos' exponent 
$\zeta=d_s/2-\theta>0$, meaning that the free-energy fluctuation
of order $L^{\theta}$ 
is a near-cancellation of much bigger energy and entropy fluctuations
of order $L^{d_s/2}$.
Within the standard droplet theory, the dynamics is characterized by
a barrier exponent $\psi$ satisfying the bounds $\theta \leq \psi \leq d-1$.
In this paper, we argue that a natural value for this barrier exponent
is $\psi=d_s/2$:
(i) for the directed polymer where $d_s=1$, this corresponds to
$\psi=1/2$ in all dimensions;
(ii) for disordered ferromagnets where $d_s=d-1$, this corresponds to
$\psi=(d-1)/2$;
(iii) for spin-glasses where interfaces have a non-trivial dimension
$d_s$ known numerically, our conjecture $\psi=d_s/2$ gives numerical 
predictions in $d=2$ and $d=3$. 
We compare these values with the available numerical results for each case,
in particular with the measure 
 $\psi \simeq 0.49$ of Kolton-Rosso-Giamarchi, 
Phys. Rev. Lett. 95, 180604 (2005)
for the non-equilibrium dynamics
of a directed elastic string.

\end{abstract}

\maketitle

\section{ Introduction  } 

The non-equilibrium dynamics of extended objects 
such as polymers or interfaces
in random media has remained very controversial over the years.
The first question is  whether the dynamics is algebraic or logarithmic, 
or equivalently whether there exists a positive barrier exponent $\psi$
as we now recall. 

\subsection{  Logarithmic dynamics with some barrier exponent $\psi>0$ } 

The activated nature of the non-equilibrium dynamics is natural within the 
 droplet scaling theory
proposed both for spin-glasses \cite{heidelberg,Fis_Hus,Fis_Hus_dyn} and for the directed polymer in a
random medium \cite{Fis_Hus_DP} (see Section \ref{droplet} below for more details).
Barriers are then expected to grow as a power-law at large scale  
\begin{eqnarray}
B(L) \sim L^{\psi}
\label{defpsi}
\end{eqnarray}
where the exponent $\psi>0$ is constant in the whole low-temperature phase $T<T_c$
and is a property of the large-scale zero-temperature fixed point.
The typical time $t_{typ}(L)$ associated to scale $L$ grows as an exponential
\begin{eqnarray}
\ln t_{typ}(L) \sim B(L) \sim L^{\psi}
\label{typtime}
\end{eqnarray}
As a consequence, the non-equilibrium dynamics starting at time $t=0$
is expected to involve only logarithmic functions of time via the characteristic
length-scale $L(t)$ associated to time $t$
\begin{eqnarray}
L(t) \sim \left( \ln t \right)^{\frac{1}{\psi}}
\label{typltime}
\end{eqnarray}

A simple one-particle one-dimensional disordered model 
where this type of activated dynamics occurs is the Sinai model
\cite{sinai}
where one particle diffuses in a random Brownian potential 
of exponent $\psi=1/2$ :
the diffusion is then logarithmic with $L(t) \sim (\ln t)^2$.
One-time and two-time properties of the non-equilibrium dynamics can be  computed at large time
via a strong-disorder  renormalization procedure that yields asymptotic exact results at large times \cite{rgsinai}.

\subsection{  Algebraic dynamics with some dynamical exponent $z$ } 

In the alternative scenario of 'algebraic' dynamics, 
barriers grow at most logarithmically with
$L$, i.e. the exponent $\psi$ of Eq. \ref{defpsi} vanishes $\psi=0$.
Time scales and length scales are then related by some dynamical exponent $z$
 \begin{eqnarray}
 t_{typ}(L)  && \sim L^{z} \\
 L(t)  && \sim t^{\frac{1}{z}}
\label{nonactivated}
\end{eqnarray}
instead of the logarithmic relations of Eqs \ref{typtime} and \ref{typltime}.
As a consequence, the aging properties of the non-equilibrium dynamics
involve ratios of times (instead of ratios of logarithms of the times ).

A simple one-particle one-dimensional disordered model 
where this type of algebraic aging occurs is 
the Bouchaud trap model \cite{trap1d} 
(see \cite{trap} for the mean-field version).

\subsection{  Debate on the phase-space structure at large scales } 

This question on the value of the barrier exponent $\psi$
 amounts more generally
 to understand the phase space
structure of polymers or interfaces in random media.
Within the droplet theory \cite{heidelberg,Fis_Hus,Fis_Hus_DP}
(see Section \ref{droplet}),
static properties are governed by low-energy excitations characterized
by the 'droplet' exponent $\theta$ and by the dimensionality $d_s$
of the surface of the excitations,
whereas the dynamics involves a priori another exponent $\psi$
satisfying the bounds $\theta \leq \psi \leq d-1$
\cite{Fis_Hus,Fis_Hus_dyn,Fis_Hus_DP}.
 Whereas the exponents $\theta$ and $d_s$ for the statics are known
either exactly or numerically in various models, 
the numerical measure of the barrier
exponent $\psi$ turns out to be much more difficult from
the point of view of computational complexity \cite{middleton}.
As a consequence, the value of $\psi$ has remain very uncertain
even numerically in many random models.
 The aim of this paper is to explain that the value $\psi=d_s/2$
 for the barrier exponent is very natural for
 disordered models that have a positive chaos exponent
 $\zeta=d_s/2-\theta>0$, and to compare with available numerical data.

\subsection{  Organization of the paper }

In Section \ref{droplet}, we recall the essential properties of
the droplet scaling theory and 
 present our general arguments for the conjecture $\psi=d_s/2$.
 We then discuss this conjecture with a comparison to
 existing numerical results for the following models :
  directed polymers in random media (Section \ref{dp}),
  disordered ferromagnets (Section \ref{rferro}),
and spin-glasses (Section \ref{sg}).
Section \ref{conclusion} contains our conclusions.

\section{ Arguments in favor of the value $\psi=d_s/2$ for the barrier exponent  } 

\label{droplet}

In this section, before explaining the conjecture 
$\psi=d_s/2$ for the barrier exponent, we need to recall the main statements
of the droplet scaling theory proposed 
both for spin-glasses \cite{heidelberg,Fis_Hus,Fis_Hus_dyn} 
and for the directed polymer in a
random medium \cite{Fis_Hus_DP}.

\subsection{ Reminder on equilibrium properties 
within the droplet theory  }

\subsubsection{ Statistics of excitations above the ground state}

At very low temperature $ T \to 0$, all observables are governed by
the statistics of low energy excitations above the ground state.
An excitation of large length $l$ costs a random energy
\begin{eqnarray}
 E_{exc}(l,T=0) \sim l^{\theta} u
\label{ground}
\end{eqnarray}
where $\theta$ is the so-called 'droplet' exponent 
\cite{heidelberg,Fis_Hus,Fis_Hus_dyn,Fis_Hus_DP}, and
where $u$ is a positive random variable distributed with some law $Q_0 (u)$
having  some finite density at the origin  $Q_0 (u=0) >0$.
A low-temperature disorder-dominated phase exists
whenever the droplet exponent $\theta$ is positive $\theta>0$.

From Eq. \ref{ground}, the probability distribution of 
large excitations $ l \gg 1$ reads within the droplet theory
\begin{eqnarray}
dl \rho_{exc} (l)  \sim \frac{ dl }{l} e^{- \beta  E_{exc}(l,T=0) } 
\sim  \frac{ dl }{l} e^{- \beta l^{\theta} u  }
\label{rhodroplet}
\end{eqnarray}
 where the factor $dl/l$ comes from the notion of independent excitations
\cite{Fis_Hus}. In particular, its average over the disorder
follows the power-law
\begin{eqnarray}
dl \overline{  \rho_{exc} (l) }  
\sim \int_0^{+\infty} du Q_0(u)  \frac{ dl }{l} e^{- \beta l^{\theta} u  }
= T Q(0) \frac{ dl }{l^{1+\theta}}
\label{rhoav}
\end{eqnarray}
Since correlation functions at large distance are directly
related to the probability of large excitations,
the low temperature phase
is very non-trivial from the point of view of correlations
lengths : the typical exponential decay of Eq \ref{rhodroplet} indicates
a finite typical correlation length $\xi_{typ}(T)$,
whereas the averaged power-law behavior of Eq. \ref{rhoav} means
that the averaged correlation length $\xi_{av}(T)$
is actually infinite in the whole low temperature phase
 $ \xi_{av}(0<T \leq T_c) =\infty$.

\subsubsection{ Low temperature phase governed by a zero-temperature fixed point}

According to the droplet theory, the whole low temperature phase $0<T<T_c$
is governed by a zero-temperature fixed point. 
However, many subtleties arise because the temperature
is actually `dangerously irrelevant'. 
The main conclusions of the droplet analysis \cite{Fis_Hus,Fis_Hus_DP}
can be summarized as follows.
The scaling of Eq \ref{ground} governs the free energy cost
of an excitation of length $l$, provided one introduces
the typical correlation length $\xi_{typ}(T)$ to rescale the length $l$
\begin{eqnarray}
 F_{exc} (l, 0<T<T_c ) = \left( \frac{l}{\xi_{typ}(T) } \right)^{\theta} u
\label{deltaF}
\end{eqnarray}
Here as before, $u$ denotes
 a positive random variable distributed with some law $Q (u)$
having  some finite density at the origin  $Q (u=0) >0$.
But this droplet free energy turns out to be
 a near cancellation of much larger energy and
entropy contributions that scale for large $l$ as \cite{Fis_Hus,Fis_Hus_DP}
\begin{eqnarray}
\label{deltaE}
E_{exc} (l, 0<T<T_c ) &&  \sim \sigma(T) l^{\frac{d_s}{2}} w  
+ e_1(T) l^{\theta}\\
T S_{exc} (l, 0<T<T_c ) &&  \sim \sigma(T) l^{\frac{d_s}{2}} w +...
\nonumber
\end{eqnarray}
where $d_s$ represents the dimension of the surface of
the excitation. The random variable $w$
of order $O(1)$ and of zero mean
is expected to be Gaussian distributed.
The argument is that the energy and entropy are dominated by small
scale contributions of random sign \cite{Fis_Hus,Fis_Hus_DP}, 
whereas the free energy
is optimized on the coarse-grained scale $\xi_{typ}(T)$.
A very important consequence of the difference in scaling of
the free-energy fluctuation of Eq \ref{deltaF} and of the 
energy-entropy fluctuations of Eq. \ref{deltaE}
is the presence of disorder and temperature chaos
in the whole low-temperature phase with the so-called chaos exponent
\cite{Ban_Bray,Fis_Hus,Fis_Hus_DP}
\begin{eqnarray}
\zeta= \frac{d_s}{2} -\theta >0
\label{zeta}
\end{eqnarray}
Note that numerically, temperature chaos is usually harder to observe
than disorder chaos (see \cite{aspel,KK} and references therein).

 For numerical simulations, it is important to stress that
the term of order $l^{ds/2}$ in Eq \ref{deltaE} is the leading term for large $l$,
but that there exists a sub-leading term of order $l^{\theta}$ to recover
the free-energy fluctuations of Eq. \ref{deltaF}.
And since the amplitude $\sigma(T)$ of the leading term
vanishes in the limit of $T=0$
(as $ \sigma(T) \propto T \times T^{1/2} = T^{3/2}$ \cite{aspel}),
whereas the amplitude $e_1(T) $ of the subleading term 
in the energy remains finite $e_1(T=0) >0$(Eq. \ref{ground}),
 one needs to simulate
sufficiently large excitations to reach the size $l$
where the leading term of Eq. \ref{deltaE} becomes much bigger
than the subleading term :
\begin{eqnarray}
\sigma(T) l^{\frac{d_s}{2}}  \gg   e_1(T) l^{\theta}
\label{largel}
\end{eqnarray}

\subsection{  Reminder on the non-equilibrium dynamical 
properties within the droplet theory }

Within the standard droplet theory \cite{heidelberg,Fis_Hus,Fis_Hus_dyn},
the non-equilibrium dynamics is governed by large-scale
barriers $B(L) \sim L^{\psi}$ where the barrier exponent $\psi$
satisfies the bounds
\begin{eqnarray}
\theta \leq \psi \leq d-1
\label{boundspsi}
\end{eqnarray}
The lower bound comes from the fact that the barrier $B(L)$ to create
a droplet excitation of size $L$ cannot be less than the free-energy cost
of the droplet (Eq \ref{deltaF}).
The upper bound comes from the expectation that the barrier
cannot have a greater exponent than the barrier $L^{d-1}$
needed to create a non-optimized excitation of surface $L^{d-1}$.
In particular, whenever there exists a low-temperature 
disorder-dominated phase with a positive droplet exponent $\theta>0$,
the barrier exponent is strictly positive $\psi \geq \theta >0$
and leads to some logarithmic dynamics (see Eq \ref{typltime}).
Since in this paper we focus on the value of the barrier exponent $\psi$,
we refer the reader to \cite{Fis_Hus,Fis_Hus_dyn,Fis_Hus_dropletsmall}
for a detailed description of other properties of the droplet dynamics.

\subsection{ Arguments in favor of the value $\psi=d_s/2$ } 

\label{argconj}

From the point of view of the dynamical exponent $\psi$,
there has been a long-standing difference between \\
(i) the directed polymer model, where 
the assumption that the barrier exponent coincides
with the droplet exponent has been made 
from the very first article that has introduced
the model \cite{Hus_Hen} (see section \ref{dp} for more details) 
\begin{eqnarray}
{\rm Usual \ Assumption \ for \ the \ directed \ polymer \ :} \ \ \  \psi = \theta
\label{falsedp}
\end{eqnarray}

(ii) spin-glasses, where it has been quickly clear that the barrier
exponent in strictly bigger than the droplet exponent
(see section \ref{sg} for more details), because they are
distinct below the lower critical dimension.
In dimension $d=1$, the exact solution \cite{heidelberg} yields
\begin{eqnarray}
{\rm 1D \ Spin-glass \ :} \ \ \  \psi=0  > \theta=-1
\label{sg1d}
\end{eqnarray}
and in dimension $d=2$
these two exponents do not have the same sign
\begin{eqnarray}
{\rm 2D \ Spin-glass \ :} \ \ \  \psi >0 > \theta
\label{sg2d}
\end{eqnarray}

The usual explanation of this difference between the two models (i) and (ii)
is that the directed polymer case
would be much more 'simple' than the spin-glass case,
that its phase space would be characterized by a single exponent $\theta$,
whereas in spin-glasses the barrier scaling is not related
to the scaling of the free-energy minima.
In this paper, we propose another scenario, based on the observation
that in any disorder system presenting a positive chaos exponent
$\zeta=d_s/2-\theta>0$ (and in particular for the directed polymer),
the description of the phase space requires at least two exponents
which are the droplet exponent $\theta$ for free-energy fluctuations
(Eq \ref{deltaF})
and the exponent $d_s/2$ that governs
 energy and entropy fluctuations (Eq \ref{deltaE}).
The assumption of Eq. \ref{falsedp} is then equivalent
to the very strong requirement
that the global free-energy optimization of order $L^{\theta}$
that results from a 
near cancellation of much bigger energy and entropy random contributions 
of order $L^{ds/2}$ is satisfied { \it  all along the dynamical trajectories }.
The alternative scenario that we propose in this paper is
that for any dynamics containing only local moves of the polymer or interface, 
 the barrier exponent $\psi$ 
 is equal to the energy-entropy
fluctuation exponent $d_s/2$ (Eq \ref{deltaE})
\begin{eqnarray}
\psi= \frac{d_s}{2} 
\label{conjecture}
\end{eqnarray}
The physical interpretation is that 
the equality $\psi=\theta$ would be possible
only via a non-local dynamics that would allow a global reorganization
of the polymer or interface at each time step,
whereas any local dynamics will see barriers that are dominated 
by small-scale contributions of random sign.

In the remaining sections, we discuss this conjecture
for various disordered models and compare with the available
numerical results.

\section{ Non-equilibrium dynamics of directed polymers in random media } 

\label{dp}

\subsection{ Reminder on the statics }

The directed polymer in a random medium (see \cite{Hal_Zha} for a review)
is a model where the various statements
of the droplet scaling theory have been successfully tested.
The exponent $\theta$ of Eq. \ref{ground}
is exactly known in one-dimension
$\theta(d=1)=1/3$ \cite{Hus_Hen_Fis,Kar,Joh}
and for the mean-field version on the Cayley tree
 $\theta(d=\infty)=0$ \cite{Der_Spo}.
In finite dimensions $d=2,3,4,5,...$, the exponent $\theta(d)$ has
been numerically measured, with values of order
 $\theta(d=2)=0.244$ and $\theta(d=3)  = 0.186$  \cite{Mar_etal,us_negtails}.
The statistics of Eq. \ref{rhoav} for the low-energy excitations
as a function of their size $l$ describes very well the numerical data in 
the regime $1 \ll l \ll L$ in dimensions $d=1,2,3$ \cite{DPexcita}.
Finally, the scaling of Eq. \ref{deltaE}
for the energy and entropy fluctuations have been numerically checked in
various dimensions in \cite{Fis_Hus_DP,Wa_Ha_Sc,us_3d}.
 Let us stress again that
the difference between free-energy fluctuations of Eq. \ref{deltaF}
and energy fluctuations of Eq. \ref{deltaE} can be seen
only for sufficiently large scale $L$ \cite{Fis_Hus_DP,Wa_Ha_Sc,us_3d}
(see the discussion before Eq \ref{largel}).

\subsection{ Discussion of the
conjecture $\psi=d_s/2=1/2$ for the barrier exponent }

To the best of our knowledge, all papers discussing the
non-equilibrium dynamics of the directed polymer
seem to have assumed the equality of Eq. \ref{falsedp}
between the barrier exponent $\psi$ and the droplet exponent $\theta$.
This assumption was first made in the very first paper
\cite{Hus_Hen} introducing the directed polymer model,
i.e. before the droplet analysis of the model \cite{Fis_Hus_DP}.
More recently, many papers consider that the 
equality of Eq. \ref{falsedp}
has been 'proven' in \cite{drossel} up to possible
 logarithmic corrections.
In our opinion, the arguments contained in \cite{drossel}
are plagued by the fact that the authors of \cite{drossel}
seem to be unaware of the crucial difference in scaling between
free-energy fluctuations and energy fluctuations (Eqs \ref{deltaF},
\ref{deltaE}). For instance, they state that
the dynamics is controlled by 'energy barriers', which have the 'same
scaling as free-energy fluctuations', because it is a 'zero-temperature
fixed point', but as recalled above, within the droplet theory,
the properties of the 'zero-temperature fixed point' are instead
the different scalings of Eqs \ref{deltaF} and \ref{deltaE}.

We have explained above in section \ref{argconj}
our general arguments in favor of the value  $\psi=d_s/2$.
For the directed polymer of dimension $d_s=1$ in a random medium
of dimension $1+d$, this corresponds to
\begin{eqnarray}
\psi_{DP}= \frac{d_s}{2} =  \frac{1}{2}
\label{conjectureDP}
\end{eqnarray}
We now compare with the available numerical results.

\subsection{ Comparison with available numerical results on the 
non-equilibrium dynamics }

\label{sec_rosso}

Although aging effects for the directed polymer 
have been fitted with algebraic time scalings by various
authors \cite{DP_alge},
the more recent work of Kolton, Rosso and Giamarchi \cite{rosso}
shows that 

(a) the growing length $L(t)$ cannot be fitted 
by a power-law $L(t) \sim t^{1/z}$ at large times,
although the short time relaxation could be fitted with some effective
exponent $z(T)$ that strongly depends on temperature. 
This could explain why the first numerical fits  \cite{DP_alge}
see apparent algebraic aging forms.

(b) the growing length $L(t)$ can be fitted with
the logarithmic form $L(t) \sim (\ln t)^{1/\psi}$ at large times,
and the value of the barrier exponent $\psi$ is asymptotically 
size and time independent as it should. 

(c) the value of $\psi$ measured in \cite{rosso}
is $\psi \simeq 0.49$ in the last three decades.
The interpretation of the authors of \cite{rosso}
that believe in the identity $\psi=\theta=1/3$,
is that barriers contain strong logarithmic corrections
$B(L) \sim L^{1/3} (\ln L)^{\mu}$.
Our interpretation is on the contrary that the measured value 
$\psi \sim 1/2$ is actually the correct one.

\section{ Non-equilibrium dynamics in disordered ferromagnets } 

\label{rferro}

\subsection{ Numerical results on coarsening in disordered ferromagnets}

The non-equilibrium dynamics in pure ferromagnets
in the low-temperature phase $T<T_c$
is well understood via the characterization
of domain coarsening \cite{bray_revue}.
In the presence of quenched disorder however, 
the large time behavior of  
the characteristic length scale $R(t)$ of the coarsening process
has remained controversial
between logarithmic behavior \cite{Hus_Hen,puri,bray_hum}
\begin{eqnarray}
R(t) \sim (\ln t)^x
\label{rtx}
\end{eqnarray}
with a universal exponent $x$ discussed below,
and power-law growth \cite{rieger_coarsening}
\begin{eqnarray}
R(t) \sim t^{\frac{1}{z(T,\epsilon)}}
\label{rtz}
\end{eqnarray}
with an exponent $z(T,\epsilon)$ that depends both on the temperature
$T$ and on the disorder strength $\epsilon$.

However, as stressed in the recent work \cite{sicilia}, 
the numerical simulations of coarsening in disordered 
ferromagnets do not really reach the large-scale large-time regime, since 
the maximal size $R_{max}$ measured is sometimes
only of order $R_{max} \sim 7$ in unit of lattice spacings 
at the end of the simulation
(see for instance Fig 4a of \cite{rieger_coarsening} 
or Fig. 1 of \cite{sicilia}).
As a consequence, the available numerical simulations on coarsening
with quenched disorder are not very conclusive
for the asymptotic regime of $R(t)$.

\subsection{ Relation between the exponent $x$ in $d=2$
and the directed polymer barrier exponent $\psi_{DP}$ }

Since the directed polymer model discussed in previous section
has been precisely introduced as a model of domain wall in
two-dimensional disordered ferromagnet \cite{Hus_Hen},
 one expects some relation between
the barrier exponent $\psi_{DP}$ of the directed polymer in $1+1$
and the exponent $x$ of Eq \ref{rtx} governing the domain growth in 
the two-dimensional disordered ferromagnet.
The first possibility would be simply  \cite{bray_hum}.
\begin{eqnarray}
x_{simple} = \frac{1}{\psi_{DP}}
\label{xsimple}
\end{eqnarray}
meaning that the dynamics is governed by the barriers
associated to the domain scale $R(t)$.
However, Huse-Henley \cite{Hus_Hen} have proposed
another scenario leading to the higher value
\begin{eqnarray}
x_{HH} = \frac{(2-\zeta)}{ \psi_{DP}}
\label{xHH}
\end{eqnarray}
where $\zeta=2/3$ is the roughness exponent of the directed polymer
in $1+1$.
The argument leading to the value of Eq \ref{xHH}
can be summarized as follows \cite{Hus_Hen,bray_hum,bray_revue}.
The relevant interfaces during the coarsening process are
not directed polymers but curved polymers with a typical curvature
radius of order $R(t)$ itself.
It can be consider as directed up to the size $l$
where the roughness $l^{\zeta}$ is of the same order 
of the curvature $l^2/R(t)$ yielding $l(t) \sim (R(t))^{1/(2-\zeta)}$.
The barriers associated to these directed parts scale as 
$(l(t))^{\psi_{DP}} \sim (R(t))^{\psi_{DP}/(2-\zeta)}$ 
leading to Eq. \ref{xHH}.

Note that these arguments usually go along with the assumption 
$\psi_{DP}=\theta_{DP}=1/3$ (see previous section on the directed polymer)
yielding the values $x_{simple}=3$ and $x_{HH}=4$ 
\cite{Hus_Hen,bray_hum,bray_revue}.
With the value $\psi_{DP}=1/2$ of our conjecture discussed
in previous section for the directed polymer, 
the values of the exponent $x$
are respectively $x_{simple}=2$ and $x_{HH}=8/3$.
In the following, we argue that within our analysis,
it is the value $x_{simple}=2$ which is natural
for disordered ferromagnets in $d=2$.

\subsection{ Conjecture $\psi=d_s/2=(d-1)/2$ for the barrier exponent  }

Within our analysis where the barrier exponent $\psi$ is governed by 
the dimensionality $d_s$ of the interface that determines
the energy and entropy fluctuations (Eq \ref{deltaE}),
we expect that in disordered ferromagnets where $d_s=d-1$,
the barrier exponent is
\begin{eqnarray}
\psi= \frac{d_s}{2}=\frac{d-1}{2}
\end{eqnarray}
irrespectively of the directed or curved nature of the interface,
since it is governed by small-scale contributions.
In particular, in domain coarsening, we expect the 'simple'
relation that generalizes Eq.
\ref{xsimple}
\begin{eqnarray}
x = \frac{1}{\psi} = \frac{2}{d-1}
\label{xsimplebis}
\end{eqnarray}
Again, as explained above, numerical data on coarsening
with disorder do not allow a precise measure of the exponent $x$
  because $R(t)$ of Eq. \ref{rtx} is never very large
in simulations (see \cite{bray_hum} for a more detailed discussions
of the results of various fits).

\section{ Non-equilibrium dynamics in spin-glasses } 

\label{sg}

Many numerical works have studied non-equilibrium properties
in spin-glasses. Here again, there is a controversy
between logarithmic dynamics (see for instance \cite{huseSG,berthier})
and algebraic dynamics (see for instance \cite{kis,kat_cam}).

\subsection{Discussion of the
conjecture $\psi=d_s/2$ for spin-glasses in $d=2$ }

In dimension $d=2$, there is no spin-glass phase because
the droplet exponent $\theta$ is negative $\theta <0$ so that $T_c=0$.
Nevertheless, it is interesting to measure the values
of the droplet exponent $\theta$ and of the fractal dimension $d_s$
of the surface of excitations above the ground state.
Recent estimates are $\theta \simeq -0.287(4)$ (see \cite{sgtheta2d}
and references therein)
and $d_s \simeq 1.274(2) $ (see \cite{melchert} and references therein).
Note that it has been recently argued that
these interfaces are described by SLE evolutions
implying some simple relation between 
$\theta$ and $d_s$ \cite{slesg}. 

Using $d_s \simeq 1.274 $ \cite{melchert},
the present conjecture $\psi=d_s/2$ for the barrier exponent
would corresponds to a numerical value of order
\begin{eqnarray}
{\rm 2D \ Spin-glass  \ :} \ \ \ \psi = \frac{d_s}{2} \simeq 0.637
\label{conjectsg2d}
\end{eqnarray}

In a recent work, Amoruso, Hartmann and Moore \cite{amoruso}
have tried to measure the barrier exponent $\psi$
of the highest barrier of systems of sizes $L \leq 40$ yielding the 
numerical bounds $0.25 < \psi < 0.54$.
The uncertainty shows the difficulty of the numerical measure of $\psi$
so that their upper bound does not seem to us sharp enough to rule out
the value of Eq \ref{conjectsg2d}.

\subsection{Discussion of the
conjecture $\psi=d_s/2$ for spin-glasses in $d=3$ }

In dimension $d=3$, 
the droplet exponent $\theta$ is positive $\theta >0$ so 
there exists a spin-glass phase with $T_c>0$.
Recent estimates for the droplet exponent $\theta$ and 
for the fractal dimension $d_s$
of the surface of excitations above the ground state
are respectively  $\theta \simeq 0.19(2) $ (see \cite{harttheta3d}
 and references therein)
and $d_s \simeq 2.6 $ (see \cite{pala} and references therein).
Using the latter, the present conjecture $\psi=d_s/2$ for the barrier exponent
would correspond to a numerical value of order
\begin{eqnarray}
{\rm 3D  \ Spin-glass  \ : } \ \ \ \psi = \frac{d_s}{2} \simeq 1.3
\label{conjectsg3d}
\end{eqnarray}
This has to be compared with the value $\psi \sim 1.0$
estimated by Berthier and Bouchaud from their
aging simulations  \cite{berthier}. Again, the precision of this
 numerical estimate does not seem sufficient to rule out
the value of Eq \ref{conjectsg3d}.
We refer the reader to  \cite{berthier} for the experimental values
of the exponent $\psi$ reported in the literature,
that varies between $0.3$ and $1.9$ \cite{berthier}.

\section{ Conclusion } 

\label{conclusion}

In this paper, we have proposed that, in disordered systems characterized by
a positive chaos exponent $\zeta=d_s/2-\theta >0$,
the large time dynamics is governed by the barrier exponent $\psi=d_s/2$.
We have explained why this value $\psi=d_s/2$ is natural
within the droplet scaling picture, where the exponent $d_s/2$ governs the energy-entropy
fluctuations  (Eq \ref{deltaE}) and is greater than the droplet exponent $\theta$
of free-energy fluctuations (Eq \ref{deltaF}).
We have then discussed our conjecture for the following models \\
(i) for the directed polymer where $d_s=1$, our conjecture
gives $\psi=1/2$  in all dimensions  ; \\
(ii) for disordered ferromagnets where $d_s=d-1$, our conjecture corresponds to
$\psi=(d-1)/2$ \\
(iii) for spin-glasses where interfaces have a non-trivial fractal dimension
$d_s$ known numerically, our conjecture $\psi=d_s/2$ gives numerical
predictions in $d=2$ and $d=3$. \\
In each case, we have compared with the available numerical data on $\psi$, in particular with 
the work of Kolton-Rosso-Giamarchi\cite{rosso}
who have measured the barrier exponent
 $\psi \simeq 0.49$ for the non-equilibrium dynamics
of a directed elastic string.
For disorder spin models, either disordered ferromagnets or spin-glasses,
the available numerical estimates of $\psi$
are not sufficiently precise to support or exclude our conjecture.

If the conjecture $\psi=d_s/2$ is correct, this means that
the numerical measure of this barrier exponent in dynamical 
simulations requires
to study samples of sizes $L$ sufficiently large,
where in the corresponding statics, the free-energy fluctuations
and the energy-entropy
fluctuations have reached their asymptotic
regimes of Eqs \ref{deltaE} and \ref{deltaF}, i.e.
one needs to be in the regime of Eq \ref{largel} for the statics.
We hope that this explicit static criterion will help to identify
 the regime where dynamic simulations 
are likely to measure the asymptotic barrier exponent $\psi$
relevant at large scales.

\end{document}